\documentclass[12pt]{article}

\usepackage[dvips]{color}
\usepackage[OT1]{fontenc}
\usepackage{anysize}
\usepackage[english]{babel}
\usepackage{multirow}
\usepackage{rotating}
\usepackage{booktabs}
\usepackage{indentfirst}
\usepackage{float}
\usepackage{latexsym}
\usepackage{geometry}
\usepackage{xcolor}

\usepackage{amsmath}
\usepackage{mathrsfs}
\usepackage{amsfonts}
\usepackage[authoryear]{natbib}
\usepackage{amsthm}
\usepackage{amssymb}
\usepackage{color}
\usepackage{lscape}
\usepackage{multirow}
\usepackage{rotating}
\usepackage{caption2}
\usepackage{subfigure}
\usepackage{graphicx}
\theoremstyle{plain}

\usepackage[plain,noend]{algorithm2e}

\def\be{\begin{equation}}
\def\ee{\end{equation}}
\def\bea{\begin{eqnarray}}
\def\eea{\end{eqnarray}}
\def\bd{\begin{displaymath}}
\def\ed{\end{displaymath}}
\def\bda{\begin{eqnarray*}}
\def\eda{\end{eqnarray*}}
\def\bsm{\begin{small}}
\def\esm{\end{small}}

\def\t0{\theta_0}

\def\ha1{\hat \beta_1}

\def\bnt{\begin{enumerate}}
\def\ent{\end{enumerate}}

\def\bsc{\begin{scriptsize}}
\def\esc{\end{scriptsize}}

\newcommand{\p}{{\rm p}}

\newtheorem{theorem}{Theorem}

\newtheorem{proposition}{Proposition}
\newtheorem{ft}{Fact}[section]
\theoremstyle{definition}

\newtheorem{condition}{Condition}

\def\boxit#1{\vbox{\hrule\hbox{\vrule\kern6pt
          \vbox{\kern6pt#1\kern6pt}\kern6pt\vrule}\hrule}}

\def\pr{\mathrm{pr}}

\makeatletter
\newcommand{\figcaption}{\def\@captype{figure}\caption}
\newcommand{\tabcaption}{\def\@captype{table}\caption}
\makeatother

\newcommand{\bi}{{\mathbf i}}

\def\SS{{\sl Statistica Sinica}}

\usepackage{fancyhdr}
\renewcommand{\baselinestretch}{1.2}

\renewcommand{\baselinestretch}{\vv}

\renewcommand{\theequation}{\arabic{equation}}
\allowdisplaybreaks

\begin{document}

\renewcommand{\baselinestretch}{1.1}\normalsize

\title{\Large \bf   Testing for unit roots based on sample autocovariances}

\author{
Jinyuan Chang$^1$\,,~~~Guanghui Cheng$^2$\,,~~~Qiwei Yao$^3$\\
\\
$^1$Southwestern University of Finance and Economics, \\
$^2$Guangzhou University, $^3$London School of Economics
}


 \date{}
\maketitle

\vspace{-0.5cm}

\begin{abstract}

We propose a new unit-root test for a stationary null hypothesis $H_0$
against a unit-root alternative $H_1$. Our approach is nonparametric as
$H_0$ only assumes that the process concerned is $I(0)$ without
specifying any parametric forms. The new test is based on the fact that
the sample autocovariance function (ACVF) converges to the finite
population ACVF for an $I(0)$ process while it diverges to infinity for a
process with unit-roots. Therefore the new test rejects $H_0$ for the
large values of the sample ACVF. To address the technical challenge `how
large is large', we split the sample and establish an appropriate normal
approximation for the null-distribution of the test statistic.  The
substantial discriminative power of the new test statistic is rooted from
the fact that it takes finite value under $H_0$ and diverges to infinity
under $H_1$. This allows us to truncate the critical values of the test
to make it with the asymptotic power one. It also alleviates the loss of
power due to the sample-splitting. 
The test is implemented in a user-friendly R-function.\end{abstract}



\baselineskip=21pt


\renewcommand{\baselinestretch}{1.2}\normalsize

\section{Introduction}\label{s1}

Models with unit-root are frequently
used for modeling nonstationary time series.
The importance of the unit-root concept stems from the fact that many economic, financial,
business and social-domain 
data exhibit segmented trend-like or random
wandering phenomena.  While the random-walk-like
behavior of stock prices was notified and recorded much earlier by, for example,
Jules Regnault, a French broker, in 1863 and then by Louis Bachelier
in his 1900 PhD thesis, the development of statistical inference for
unit-roots only started in late 1970s.
Nevertheless the literature on unit-root tests by now is immense and
diverse. We only state a selection of some important developments below,
which naturally leads to a new test to be presented in this paper.

 The Dickey-Fuller tests \citep{DF1979, DF1981} dealt
with Gaussian random walks with independent error terms. Effort to
relax the condition of independent Gaussian errors leads to, among
others, 
the augmented Dickey-Fuller (ADF) tests
\citep{Said1984, Elliot1996} 
 which replace the error term by an autoregressive process, the
Phillips-Perron test \citep{Phillips1987, Phillips1988} which
estimates the long-run variance of the error process nonparametrically.
The ADF tests are further extended for dealing with structural breaks
in trend \citep{Zivot1992}, long memory processes \citep{Robinson1994},
seasonal unit roots \citep{Chan1988,Hylleberg1990}, bootstrap unit-root
tests \citep{Paparoditis2005}, nonstationary volatility \citep{Cavaliere2007},
 panel data \citep{Pesaran2007},
and local stationary processes \citep{RhoShao2019}. We refer to
survey papers \cite{Stock1994} and \cite{PhillipsXiao1998}, and
monographs \cite{Hatanaka1996} and \cite{Maddala1998} for further
references.

The Dickey-Fuller tests and their variants are based on the regression of a
time series on its first lag in which
the existence of unit root is postulated as a null hypothesis in the form
of the regression coefficient being equal to one. This null hypothesis is tested
against a stationary alternative that  the regression coefficient is smaller than one. This setting leads to innate indecisive
inference for ascertaining the existence of unit-roots, as a statistical
test is incapable in accepting null hypothesis. To make the assertion
of unit-roots on a firmer ground, \cite{kpss1992} adopted a different
approach: the proposed KPSS test considers a stationary null hypothesis
against a unit-root alternative.  It is based on a plausible
representation for possible nonstationary time series in which
a unit-root is represented as an additive random walk component.
Then under null the variance
of the random walk component is zero.  The KPSS test is the one-sided
Lagrange multiplier test for testing the variance to be zero against greater than zero.

In spite of the many exciting developments stated above, testing for the
existence of unit roots remains as a  challenge
in time series analysis, as most available methods suffer from
the lack of accurate size control and poor power.
In this paper we propose a new test, based on a radically different idea
from the existing approaches. Our setting is similar in spirit to the KPSS test
as we test for stationary null hypothesis $H_0$ again a unit-root alternative $H_1$.
However our approach is nonparametric as $H_0$ only assumes
that the process concerned is $I(0)$ without specifying any parametric forms. The new test is based on the simple fact that under $H_0$
the sample autocovariance function (ACVF) converges to the
finite population ACVF while under $H_1$
it diverges to infinity. Therefore we can reject $H_0$ for large (absolute) values of the sample ACVF. To address the technical
challenge `how large is large', we split the sample and establish an appropriate
normal approximation for the null-distribution of the test statistic. Note that
our sample ACVF based test statistic offers substantial discriminative power as it
takes finite value under $H_0$ or diverges to infinity
under $H_1$.
This allows us to truncate the critical values determined by the normal approximation to make the test with the asymptotic power one.
Furthermore, it also alleviates the loss of power due to the sample-splitting
as it outperforms the KPSS test in the power comparison in simulation.
Another advantage of the new method is that it has a remarkable discriminate
power to tell the difference between, for example, a random walk and an
AR(1) with the autoregressive coefficient close to (but still smaller than) one, for
which most the available unit-root tests, including the KPSS method, suffer from
weak discriminate power. Admittedly the new test is technically sophisticated, which, we argue, is inevitable in order to gain improvement over the existing methods.
Nevertheless to make
it user-friendly, we have developed an R-function {\tt ur.test} in the
package {\tt HDTSA} which implements the test in an automatic manner.



\section{Main results}\label{secc2}

\subsection{A power-one test}\label{sec21}

A time series
$\{Y_t\}$ is said to be $I(0)$, denoted by $Y_t\sim I(0)$, if
$
E(Y_t)\equiv\mu$, $E(Y_t^2)<\infty$, $\gamma(k)\equiv\textrm{Cov}(Y_{t+k},Y_t)$, 
and $\sum_{k=0}^\infty|\gamma(k)|<\infty$.
 Let $\nabla Y_t=Y_t-Y_{t-1}$, $\nabla^0 Y_t=Y_t$ and $\nabla^dY_t=\nabla(\nabla^{d-1}Y_t)$
for any integer $d\ge1$. $\{Y_t\}$ is said to be $I(d)$, denoted by
$Y_t\sim I(d)$, 
 if $\{\nabla^dY_t\}$ is $I(0)$ and $\{\nabla^{d-1}Y_t\}$ is not $I(0)$.
An $I(d)$ process is also called a unit-root process with the integration order $d$.
With the observations $\{Y_t\}_{t=1}^n$, we are interested in testing the hypotheses
\begin{equation}\label{a1}
H_0: Y_t\sim I{(0)}~~~{\rm versus}~~~H_1: Y_t\sim I(d)~\textrm{for some integer}~d\geq1\,.
\end{equation}
Write $\bar{Y}=n^{-1}\sum_{t=1}^nY_t$ and denote the sample ACVF at lag $k$ by
$
\hat{\gamma}(k)=n^{-1}\sum_{t=1}^{n-k}(Y_{t+k}-\bar{Y})({Y_t-\bar{Y}})$,
which is consistent estimator for $\gamma(k)$ under $H_0$. 
Proposition \ref{pn:alt} indicates that
$\hat\gamma(k)$ diverges to infinity under $H_1$. 
Thus we can reject $H_0$ for large values of $|\hat{\gamma}(k)|$. When $Y_t \sim I(d)$, the
Wold's decomposition for the purely non-deterministic $I(0)$ process admits 
\begin{align}\label{eq:wolddec}
\nabla^dY_t=\mu_d+\sum_{j=0}^\infty \psi_j\epsilon_{t-j}
\end{align}
where $\mu_d=E(\nabla^dY_t)$ is a constant, $\psi_0=1$,
and $\{\epsilon_t\}$ is a white noise sequence. 


\begin{proposition}\label{pn:alt}
Let $Y_t$ satisfy {\rm(\ref{eq:wolddec})} with independent $\epsilon_t\sim(0,\sigma_{\epsilon}^2)$ and
$\sum_{j=1}^\infty j|\psi_j|<\infty$. Write $a=\sum_{j=0}^\infty\psi_j$ and
$V_{d-1}(t)=F_{d-1}(t)-\int_0^1F_{d-1}(t)\,{\rm d}t$ with the
scalar multi-fold
 integrated Brownian motion $F_{d-1}(t)$ defined recursively as
$F_j(t)=\int_0^tF_{j-1}(x)\,{\rm d}x$ for any $j\geq1$ and the  standard Brownian motion $F_0(t)$.
For any given integer $k\geq 0$, as $n\rightarrow\infty$, it holds that {\rm(i)}
$
n^{-(2d-1)}\hat{\gamma}(k)\rightarrow a^2\sigma_{\epsilon}^2\int_0^1V_{d-1}^2(t)\,{\rm d}t
$ in distribution if $\mu_d=0$, and
{\rm(ii)}
$
n^{-2d}\hat{\gamma}(k)\rightarrow\phi_{d,k}\mu_d^2
$ in probability if $\mu_d\neq 0$,
where $\phi_{d,k}>0$ is a bounded constant only depending on $d$ and $k$.
\end{proposition}


By Proposition \ref{pn:alt}, one may consider to reject $H_0$ for the
large values of 
$T_{{\rm naive}}=\sum_{k=0}^{K_0}|\hat{\gamma}(k)|^2$ with a prescribed integer $K_0 \ge 0$, as $T_{{\rm naive}}$ converges to $\sum_{k=0}^{K_0}|{\gamma}(k)|^2<\infty$ under $H_0$. Unfortunately, there are two
obstacles preventing using $T_{{\rm naive}}$: (i) to determine the critical values one has to derive the null-distribution
of $a_n\{T_{{\rm naive}}-\sum_{k=0}^{K_0}|{\gamma}(k)|^2\}$ with 
some $a_n\rightarrow\infty$, (ii) one needs a consistent estimator for $\sum_{k=0}^{K_0}|{\gamma}(k)|^2$
under $H_0$, which is not readily available as we do not know
if $H_0$ holds or not in practice. To overcome these two obstacles, we implement the idea of `data splitting'.
Let $N=\lfloor n/2\rfloor$. Define
$
\hat{\gamma}_1(k)=N^{-1}\sum_{t=1}^{N-k}(Y_{t+k}-\bar{Y})(Y_t-\bar{Y})$ and $
\hat{\gamma}_{2}(k)=N^{-1}\sum_{t=N+1}^{2N-k}(Y_{t+k}-\bar{Y})(Y_t-\bar{Y})$.
The test statistic for \eqref{a1} is defined as
\begin{equation*} \label{Tn}
T_n=\sum_{k=0}^{K_0}|\hat{\gamma}_2(k)|^2\,,
\end{equation*}
where $K_0 \ge 0$ is a prescribed integer which controls the amount of information
from different time lags to be used. Although our theory allows $K_0$ diverging with
sample size $n$, the simulation results reported in \S 3
indicate that the finite sample
performance of the test is robust with respect to the different
values of $K_0$ and it works well even
with small $K_0$.

Formally we
reject $H_0$ at
the significance level $\phi \in (0, 1)$ if
$
T_n>\textrm{cv}_\phi$, where $\textrm{cv}_\phi$ is the critical value satisfying
$\pr_{H_0}(T_n>\textrm{cv}_\phi)\rightarrow\phi$. 
As we will see in (\ref{eq:norcv}), $\{\hat{\gamma}_1(k)\}_{k=0}^{K_0}$ are 
used to determine the critical value $\textrm{cv}_\phi$.
One obvious concern for splitting the sample into two halves is the loss in testing
power. However the fact that $T_n$ takes finite values under $H_0$ and it
diverges to infinity under $H_1$ implies
that $T_n$ has a strong discriminant power to tell apart $H_1$ from $H_0$, which
is enough to sustain the adequate power in comparison to that of, for example,
the KPSS test. Our simulation results indicate that the
sample-splitting works well even for sample size $n=80$. Under $H_0$, write
$
y_{t,k}=2\{(Y_t-\mu)(Y_{t+k}-\mu)-\gamma(k)\}{\rm sgn}(k+t-N-1/2)$.
For $\ell\geq1$, define $B_\ell^2=E\{(\sum_{t=1}^\ell Q_t)^2\}$ with $Q_t=\sum_{k=0}^{K_0}\xi_{t,k}$ and  $\xi_{t,k}=2y_{t,k}\gamma(k)$. 
The following regularity conditions are now in order. 
See the supplementary material for the discussion of their validity. 


\begin{condition}\label{as:moment}
Under $H_0$, $\max_{1\leq t \leq n}E(|Y_t|^{2s_1})\leq c_1$ for two constants $s_1\in(2,3]$ and $c_1>0$.
\end{condition}


\begin{condition}\label{as:alphamixing}
Under $H_0$, $\{Y_t\}$ is $\alpha$-mixing with $\alpha(\tau)=\sup_{t}\sup_{A \in {\mathcal F}_{-\infty}^t, B \in {\mathcal F}_{t+\tau}^{\infty}}|\pr(AB)-\pr(A)\pr(B)|\leq c_2\tau^{-\beta_1}$ for any $\tau\geq1$, where ${\mathcal F}_{-\infty}^t$ and $\mathcal{F}_{t+\tau}^\infty$ denote the $\sigma$-fields generated by $\{Y_u\}_{u\leq t}$ and $\{Y_u\}_{u\geq t+\tau}$, respectively, $c_2>0$ and $\beta_1>2(s_1-1)s_1/(s_1-2)^2$ are two constants with $s_1$ specified in Condition \ref{as:moment}. 
\end{condition}


\begin{condition}\label{as:varb}
Under $H_0$, there is  a constant $c_3>0$ such that $B_\ell^2\geq c_3\ell$ for any $\ell\geq1$.
\end{condition}





\begin{theorem}\label{th:1}
Let $H_0$ hold with Conditions {\rm\ref{as:moment}}--{\rm\ref{as:varb}} being satisfied, and
$K_0=o\{n^{\xi(\beta,s_1)}\}$ with
$
\xi(\beta,s_1)=\min[(s_1-2)/(4s_1),(\beta-1)(s_1-2)/\{(2\beta+2)s_1\}]
$, 
where $s_1$ and $ \beta_1$ are specified, respectively, in Conditions {\rm\ref{as:moment}} and {\rm\ref{as:alphamixing}}, and $\beta=\beta_1(s_1-2)^2/\{2s_1(s_1-1)\}$. Then, as $n\rightarrow \infty$,
\begin{align*}
\sup_{u>0}\bigg|\pr\bigg\{\sqrt{n} T_n > u+\sqrt{n}\sum_{k=0}^{K_0}|\hat{\gamma}_1(k)|^2\bigg\}-1+\Phi\bigg(\frac{2Nu}
{B_{2N-K_0}\sqrt{n}}\bigg)\bigg|\rightarrow0\,.
\end{align*}
\end{theorem}


One may select the critical value 
as ${\rm cv}_{\phi,{\rm naive}}=z_{1-\phi}\hat{B}_{2N-K_0}/(2N)+\sum_{k=0}^{K_0}|\hat{\gamma}_1(k)|^2$,
where $z_{1-\phi}$ is the $(1-\phi)$-quantile of $\mathcal{N}(0,1)$ and $\hat{B}_{2N-K_0}$ is an estimate of $B_{2N-K_0}$ satisfying the condition 
 $\hat{B}_{2N-K_0}/B_{2N-K_0}\rightarrow1$ in probability under $H_0$
(see \S \ref{se:longrunest} and Theorem \ref{tm:2}),
as then the rejection probability of the test under $H_0$ converges to
$\phi$. 
Unfortunately 
$\sum_{k=0}^{K_0}|\hat{\gamma}_1(k)|^2$ diverges to infinity 
under $H_1$. This causes substantial power loss.
To rectify
this defect, we apply here the truncation idea as in \S 2.3 of \cite{Changetal2017}.
More precisely we set the critical value as
\begin{equation}\label{eq:norcv}
\textrm{cv}_\phi={\rm cv}_{\phi,{\rm naive}}\cdot I(\mathcal{T})+
\kappa_n\cdot I(\mathcal{T}^c)\,,
\end{equation}
where $\kappa_n=0.1\times\log N$ with $N=\lfloor n/2\rfloor$, and the event $\mathcal{T}$
satisfies conditions $\pr_{H_0}(\mathcal{T}
)\rightarrow1$ and $\pr_{H_1}(\mathcal{T}^c
)\rightarrow1$ as $n\rightarrow\infty$. 
Note that
 $\pr_{H_0}(\textrm{cv}_\phi=
{\rm cv}_{\phi,{\rm naive}})\rightarrow1$ and $\pr_{H_1}(\textrm{cv}_\phi=\kappa_n)\rightarrow1$. The former one makes the rejection probability of the proposed test under $H_0$ converges to the nominal level $\phi$. 
Proposition \ref{pn:alt} shows that $\pr_{H_1}\{|\hat{\gamma}_2(0)|^2>\kappa_n\}\rightarrow1$. Due to $T_n\geq |\hat{\gamma}_2(0)|^2$, we have $\pr_{H_1}(T_n>\kappa_n)\rightarrow1$, which entails that
the proposed test has power one asymptotically. We will state in \S \ref{se:cv} how to
specify a qualified event $\mathcal{T}$. 


\begin{theorem}\label{tm:size}
 Let ${\rm cv}_\phi$ be defined by {\rm(\ref{eq:norcv})} with ${\mathcal T}$ satisfying $\pr_{H_0}(\mathcal{T}
)\rightarrow1$ and $\pr_{H_1}(\mathcal{T}^c
)\rightarrow1$, and $\hat{B}_{2N-K_0}/B_{2N-K_0} \rightarrow1$ in probability under $H_0$,
 as $n\rightarrow\infty$. Then it holds that
  {\rm(i)} $
\pr_{H_0}(T_n>{\rm cv}_\phi)\rightarrow\phi$ if the conditions of Theorem {\rm\ref{th:1}} hold, and {\rm(ii)} $
\pr_{H_1}(T_n>{\rm cv}_\phi)\rightarrow1$ if
 $Y_t$ satisfies {\rm(\ref{eq:wolddec})} with independent $\epsilon_t\sim (0,\sigma_{\epsilon}^2)$ and
$\sum_{j=1}^\infty j|\psi_j|<\infty$.

\end{theorem}





\subsection{Determining the event $\mathcal{T}$ in \eqref{eq:norcv}}\label{se:cv}

The critical value cv$_\phi$ defined in (\ref{eq:norcv}) depends
on event $\mathcal{T}$
critically. Let $X_t=\nabla
Y_t$ and
$
\hat{\gamma}_x(k)=(n-1)^{-1}\sum_{t=2}^{n-k}(X_{t+k}-\bar{X})(X_t-\bar{X})$
for $k\geq0$, where $\bar{X}=(n-1)^{-1}\sum_{t=2}^nX_t$. 
 To avoid the effect of the innovation variance $\sigma_\epsilon^2$, we consider the ratio
$
R=\{\hat{\gamma}(0)+\hat{\gamma}(1)\}/\{\hat{\gamma}_x(0)+\hat{\gamma}_x(1)\}$.
Notice that $
R=O_{\p}(1)$ under $H_0$, and
$
\pr_{H_1}(R\geq C_*N^{3/5})\rightarrow1
$ for any fixed constant $C_*>0$. We define $\mathcal{T}$ in (\ref{eq:norcv}) as follows:
\begin{equation}\label{eq:Tevent}
\mathcal{T}=\{R< C_*N^{3/5}\}\,.
\end{equation}
To use $\mathcal{T}$ with  finite samples, $C_*$
must be specified according to the underlying process.


\begin{proposition}\label{pn:ratio}
Let $Y_t\sim I(1) $ satisfy {\rm(\ref{eq:wolddec})} with independent $\epsilon_t\sim(0,\sigma_{\epsilon}^2)$ and
$\sum_{j=1}^\infty j|\psi_j|<\infty$. Write $\eta=\sum_{j=0}^{\infty}\psi_j^2+\sum_{j=0}^{\infty}\psi_j\psi_{j+1}$. As $n\rightarrow\infty$, we have
{\rm (i)}
${n}^{-1}R
\rightarrow 2a^2\eta^{-1}\int_0^1V_{0}^2(t)\,{\rm d}t$ in distribution if $\mu_1=0$, 
where $a$ and $V_0(t)$ are defined in Proposition \ref{pn:alt}, and {\rm (ii)}
$n^{-2}R\rightarrow 6^{-1}\sigma_{\epsilon}^{-2}\eta^{-1}\mu_1^2$ in probability if $\mu_1\neq 0$.
\end{proposition}


Proposition \ref{pn:ratio} shows that
$R$
with $\mu_1\neq 0$ diverges faster than that with
$\mu_1=0$. Thus for any given $C_*>0$ the requirement
$\pr_{H_1}(\mathcal{T}^c)\rightarrow1$
is satisfied more readily with $\mu_1\neq 0$. Hence we focus on the cases with $\mu_1=0$ only. Recall $X_t=\nabla Y_t=\mu_1+\sum_{j=0}^\infty\psi_j\epsilon_{t-j}$.
Then
$
a^2\eta^{-1}=\lambda^{-1}(1+\rho)^{-1}$,
where
$
\rho=(\sum_{j=0}^\infty\psi_j^2)^{-1}\sum_{j=0}^\infty\psi_j\psi_{j+1}$
is the first order autocorrelation coefficient,
and $\lambda={\sigma_S^2}/{\sigma_L^2}$ with the short-run variance
$
\sigma_S^2=\sigma_{\epsilon}^2\sum_{j=0}^{\infty}\psi_j^2$ and the long-run variance $\sigma_L^2=\sigma_{\epsilon}^2(\sum_{j=0}^{\infty}\psi_j)^2$.
Write 
$
\hat{\sigma}_S^2=\hat{\gamma}_x(0)
$. Applying the estimation method for the long-run variance suggested in \S
\ref{se:longrunest}, we can obtain $\hat{\sigma}_L^2$, 
the kernel-type estimate of $\sigma_L^2$,  based on  
$\{X_t-\bar{X}\}_{t=2}^n$. Then we can estimate $\lambda$ and $\rho$ by
$
\hat{\lambda}=\hat{\sigma}_S^2/\hat{\sigma}_L^2$ and $\hat{\rho}=\hat{\gamma}_x(1)/\hat{\gamma}_x(0)$. 
As $E\{\int_0^1V_0^2(t)\,{\rm d}t\}=1/6$, we now specify the
model-dependent constant $C_*$ in \eqref{eq:Tevent} as
\begin{equation}\label{eq:Cstar}
C_*=2c_{\kappa}/\{\hat{\lambda}(1+\hat{\rho})\}
\end{equation}
for some constant $c_\kappa>1/6$.
Our extensive simulation results indicate that
this specification of $C_*$ with $c_\kappa
\in[0.45,0.65]$ works well across a variety of models. 

Though the above specification was derived
for $Y_t \sim I(1)$, our simulation results indicate that it also
works well for $I(2)$ processes. Note that testing
$I(0)$ against $I(d)$ with $d>1$ is easier than that with $d=1$, as
the autocovariances  are of the order at least $n^{2d-1}$ for $I(d)$ processes.
So the difference between the values of
$T_n$ under $H_1$ and those under $H_0$
increases as $d$ increases.

\subsection{Estimation of $B_{2N-K_0}^2$}\label{se:longrunest}

Write $m=2N-K_0$. Recall $Q_t=\sum_{k=0}^{K_0}\xi_{t,k}$ with $\xi_{t,k}=2y_{t,k}\gamma(k)$. Let $V_m$ be the long-run variance of the sequence $\{Q_t\}_{t=1}^m$. We then have 
$
B_{2N-K_0}^2=mV_m$. 
Define $
\tilde{Q}_t=\sum_{k=0}^{K_0}\tilde{\xi}_{t,k}$ with $\tilde{\xi}_{t,k}=2\tilde{y}_{t,k}\hat{\gamma}(k)$, where   $\tilde{y}_{t,k}=2\{(Y_t-\bar{Y})(Y_{t+k}-\bar{Y})-\hat{\gamma}(k)\}{\rm sgn}(k+t-N-1/2)$. Write $\tilde{G}_j=m^{-1}\sum_{t=j+1}^m\tilde{Q}_t\tilde{Q}_{t-j}$ if $j\geq0$ and $\tilde{G}_j=m^{-1}\sum_{t=-j+1}^m\tilde{Q}_{t+j}\tilde{Q}_t$ otherwise.
We can estimate $V_m$ by 
$
\tilde{V}_m=\sum_{j=-m+1}^{m-1}\mathcal{K}(j/b_m)\tilde{G}_j$ with a kernel $\mathcal{K}(\cdot)$ and 
bandwidth $b_m$.  Let
\begin{equation*}\label{eq:hatB}
\hat{B}_{2N-K_0}=(m\tilde{V}_m)^{1/2}\,.
\end{equation*}
\cite{Andrews1991} found that the Quadratic Spectral kernel
 is optimal for such estimation. 
We suggest using this
kernel in practice by calling function {\tt lrvar} from the R-package
{\tt sandwich} with the default bandwidth specified in the function. To state the required asymptotic property for $\hat{B}_{2N-K_0}$ with general kernels, we need following regularity conditions.


\begin{condition}\label{as:kernel}
The kernel function $\mathcal{K}(\cdot):\mathbb{R}\rightarrow[-1,1]$ is continuously differentiable on $\mathbb{R}$ and satisfies conditions: (i) $\mathcal{K}(0)=1$, (ii) $\mathcal{K}(x)=\mathcal{K}(-x)$ for any $x\in\mathbb{R}$, and (iii) $\int_{-\infty}^{\infty}|\mathcal{K}(x)|\,{\rm d}x<\infty$. Let $K_*=K_0+2$ satisfying $K_*^{13}\log K_*=o(n^{1-2/s_2})$ with $s_2$ specified in Condition \ref{as:newmomalph}. The bandwidth $b_m\rightarrow\infty$ as $n\rightarrow\infty$ satisfies $b_m=o\{n^{1/2-1/s_2}(K_*^5\log K_*)^{-1/2}\}$ and
$K_*^4=o(b_m)$.
\end{condition}


\begin{condition}\label{as:newmomalph}
Under $H_0$, $\max_{1\leq t \leq n}E(|Y_t|^{2s_2})\leq c_4$ for two constants $s_2>4$ and $c_4>0$, and the $\alpha$-mixing coefficients $\{\alpha(\tau)\}_{\tau\geq1}$ satisfy $\alpha(\tau)\leq c_5\tau^{-\beta_2}$ for two constants $c_5>0$ and $\beta_2>\max\{2s_2/(s_2-2), s_2/(s_2-4)\}$, where $\alpha(\tau)$ is defined in Condition \ref{as:alphamixing}.
\end{condition}


\begin{theorem}\label{tm:2}
Let Conditions {\rm\ref{as:kernel}} and {\rm\ref{as:newmomalph}} hold.
Then as $n\rightarrow\infty$,
$\hat{B}_{2N-K_0}/B_{2N-K_0}\rightarrow1$ in probability
under $H_0$. 
\end{theorem}

\subsection{Implementation of the test} \label{sec_impl}

Based on \S \ref{se:cv} and \S \ref{se:longrunest}, Algorithm 1
outlines the steps to perform our test which includes two tuning parameters.
The algorithm is implemented in an
R-function {\tt ur.test} in the package {\tt HDTSA} 
available at CRAN. 
To perform the test using function {\tt ur.test}, one merely needs to input
time series $\{Y_t\}_{t=1}^n$ and nominal level $\phi$.
The package sets the default value $c_\kappa=0.55$ and returns
the five testing results for $K_0=0, 1, \ldots, 4$ respectively.
One can also set $(c_\kappa,K_0)$ subjectively.
We recommend to use $c_\kappa \in [0.45, 0.65]$ and $K_0\in\{0,1,2,3,4\}$.


\begin{algorithm}[h]
\vspace*{-15pt}
\caption{Sample ACVF-based unit-root test.} \label{al1}
\begin{tabbing}
Input: Time series $\{Y_t\}_{t=1}^n$, nominal level $\phi$, and two
(optional) tuning parameters $(c_\kappa,K_0)$.\\

    Step 1. Compute $\hat{\gamma}(k)$,
$\hat{\gamma}_1(k)$, $\hat{\gamma}_2(k)$ and $\hat{\gamma}_x(k)$. Put
$\hat{\rho}=\hat{\gamma}_x(1)/\hat{\gamma}_x(0)$.\\

    Step 2. Call function \verb"lrvar" from the R-package \verb"sandwich" (with
the default bandwidth \\
 ~~~~~~~~~~~~in the function) to compute
the long-run variances of $\{\tilde{Q}_t\}$ and $\{X_t\}$, denoted by\\

 ~~~~~~~~~~~~$\tilde{V}_{2N-K_0}$ and $\hat{\sigma}_L^2$, respectively, where
$\tilde{Q}_t$ is defined in \S \ref{se:longrunest}. Put $\hat{\lambda}=\hat{\gamma}_x(0)/\hat{\sigma}_L^2$.\\

 Step 3. Calculate the test statistic $T_n=\sum_{k=0}^{K_0}|\hat{\gamma}_2(k)|^2$, and the critical value ${\rm cv}_\phi$ as in \eqref{eq:norcv}\\
 ~~~~~~~~~~~~with $\hat{B}_{2N-K_0}=(2N-K_0)^{1/2}\tilde{V}_{2N-K_0}^{1/2}$ and $\mathcal{T}$ given in \eqref{eq:Tevent} for $C_*$ specified in \eqref{eq:Cstar}.\\

 Step 4. Reject $H_0$ if $T_n>{\rm cv}_\phi$.
\end{tabbing}
\end{algorithm}

\vspace{-15pt}

 To illustrate the robustness 
with respect to the choice of $(c_\kappa,K_0)$, 
we apply our test to
 the 14 US annual economic time series 
\citep{np82} that were often used for testing unit-roots in the literature;
leading to the exactly same results with 
$c_\kappa \in\{0.45, 0.55,0.65\}$ and $K_0\in\{0,1,2, 3,4\}$ for each of
the 14 time series.
 See the details in the supplementary material.

\section{Simulation study}
\label{sec_simu}
We illustrate the finite sample properties of our test $T_n$ by simulation with  $K_0\in\{0,1,2,3,4\}$ and
$c_\kappa\in\{0.45, 0.55, 0.65\}$. We also consider $T_n$ with the untruncated critical value ${\rm cv}_{\phi,{\rm naive}}$, i.e. $c_\kappa=\infty$ in (\ref{eq:Cstar}).
\cite{HualderRobinson2011} proposed the pseudo MLE $\hat{d}$ for the integration order $d$ in the ARFIMA models that can be used to construct a $t$-statistic $\hat{d}/{\rm sd}(\hat{d})$ for $H_0: d=0$ versus $H_1:d\geq1$. We call it HR test that rejects $H_0$ if $\hat{d}/{\rm sd}(\hat{d})>z_{1-\phi}$, where $z_{1-\phi}$ is the $(1-\phi)$-quantile of $\mathcal{N}(0,1)$. For comparison, we also include the KPSS test \citep{kpss1992} and the HR test in our experiments.  We set $N=40, 70, 100$ and repeat each setting 2000 times. To examine the rejection probability of the tests under $H_0$, we consider three models: 

Model 1. $Y_t=\rho Y_{t-1}+\epsilon_t$\,.

Model 2. $Y_t=\epsilon_t+\phi_1\epsilon_{t-1}+\phi_2\epsilon_{t-2}$\,.

Model 3. $Y_t-\rho_1Y_{t-1}-\rho_2
Y_{t-2}=\epsilon_t+ 0.5 \epsilon_{t-1}+ 0.3 \epsilon_{t-2}$\,.

\noindent To examine the rejection probability of the tests under $H_1$, we consider the following models:

Model 4. $\nabla Y_t=Z_t$, $Z_t=\rho Z_{t-1}+\epsilon_t$\,. 

Model 5. $\nabla Y_t=Z_t$, $Z_t=\epsilon_t+\phi_1\epsilon_{t}+\phi_2\epsilon_{t-1}$\,. 

Model 6. $\nabla Y_t=Z_t$, $Z_t-\rho_1Z_{t-1}-\rho_2 Z_{t-2}=\epsilon_t+0.5\epsilon_{t}+0.3\epsilon_{t-1}$\,. 

Model 7. $\nabla^2 Y_t=Z_t$, $Z_t=\epsilon_t+\phi_1\epsilon_{t}+\phi_2\epsilon_{t-1}$\,.

\noindent Unless specified otherwise, we always assume that $\epsilon_t\sim\mathcal{N}(0, \sigma_{\epsilon}^2)$ independently with
$\sigma_\epsilon^2=1$ or 2, and set the nominal level $\phi=5\%$. 
The results with different $(c_\kappa,K_0)$ are similar;
indicating once again that our test is robust with respect to
the choice of $(c_\kappa,K_0)$. We only list the results with $K_0=0$ and $\sigma_{\epsilon}^2=1$ in Table \ref{tab1}, and report other results in the supplementary
material.
We also consider the cases $\epsilon_t\sim t(2)$ and $t(5)$, and
report the results in the supplementary material.

Overall the rejection probabilities of our test under $H_0$ are close to the nominal level $\phi=5\%$ especially with large $n$ ($N=100$). 
The performance of our test
is stable across different models with different parameters, different $K_0$
and different innovation distributions, while that of the KPSS test and the HR test
vary and are adequate only for some settings. Table \ref{tab1} indicates that our test works well for Model 1 with both positive and negative $\rho$, while the KPSS test and the HR test perform poorly
when $\rho<0$, and even worse  when $\rho>0$. The
KPSS test and the HR test completely fail when
$\rho=0.9$, as the rejection probabilities are at least 46.7\%. This is due to the fact
that when $\rho$ is close to 1, the KPSS test and the HR test have difficulties in
distinguishing it from 1 which is unit-root. See also Table 3 of \cite{kpss1992}.
Our test does not suffer from this closeness to 1, as for which the order
of the magnitude
of ACVF matters. Our test and the KPSS test work well for Model 2 while the HR test is too conservative. For Model 3, the rejection probabilities of our test and the HR test are close to $5\%$ while the KPSS test does not work as its rejection probabilities
range from 16.6\% to 26.2\%. Our test with $c_\kappa=\infty$ has no power which shows that the truncation step for the critical value in \eqref{eq:norcv} is necessary. 
The KPSS test has impressive power due to the fact that it has a tendency to overestimate the rejection probability under $H_0$, leading to inflated power. Nevertheless our test
displays greater power in most cases in comparison to the KPSS test. The HR test has good power in Models 4 and 5 while it performs poorly in Model 6. 
The power one property of our test 
is observable in the simulation as the rejection probability tends to 1 when $N$ increases. Comparing the results of Models 5 and 7, we found that our tests show off the power one property
more distinctly as our test statistic has
more discriminate power between $I(2)$ and $I(0)$ than that between
$I(1)$ and $I(0)$. 


\begin{table}
\centering
{\scriptsize
\caption{The rejection probabilities $(\%)$ of the proposed test $T_n$ with $K_0=0$ and $c_\kappa=0.45, 0.55$, $ 0.65, \infty$, the KPSS
test and the HR test. 
The
nominal level is $5\%$.}
\label{tab1}
\setlength\tabcolsep{5pt}
\begin{tabular}{cccccccccccccccc}
 \multicolumn{8}{c}{Model 1}& \multicolumn{8}{c}{Model 4}\\
 $\rho$ &$N$&{$\infty$} &
$0.45$&{ $0.55$} &{$0.65$}& {KPSS}& {HR} &$\rho$ & $N$&{$\infty$} &
$0.45$&{ $0.55$} &{$0.65$}& {KPSS} & {HR}\\
  0.5& 40 &6.0&6.0&6.0&6.0&10.4&5.7 & 0.5 &40 &11.7&94.2&88.4&84.0&84.2&96.4 \\
            &70  &6.9&6.9&6.9&6.9&10.1&7.0 &&70&11.7&96.5&92.9&88.4&90.9& 99.8  \\
            &100 &6.1&6.1&6.1&6.1&10.2&8.4 &&100&11.3&98.0&95.5&92.2&95.5& 100\\
         0.9& 40 &7.2&41.9&30.0&20.3&51.2&46.8 &0.9&40&13.1&99.2&97.3&94.6&91.1  &98.9 \\
          &70&  7.8&23.7&14.6&10.4&46.7& 58.8 &&70&14.8&99.8&99.1&97.9&95.3  &100 \\
          &100& 8.5&12.7&9.4&8.6&49.2& 61.1 &&100&16.4&99.9&99.5&99.1&97.2 & 100 \\
   $-$0.5& 40 &7.4&7.4&7.4&7.4&1.8 & 0.1 &$-$0.5&40&5.6&82.2&75.1&67.6&81.5 &99.7  \\
          &70& 6.9&6.9&6.9&6.9&2.5& 0.2 &&70&6.3&92.1&86.1&80.0&90.1& 100 \\
          & 100& 6.4&6.4&6.4&6.4&1.8& 0.3 && 100&5.8&94.2&89.5&85.2&94.5&100 \\

\multicolumn{8}{c}{Model 2}& \multicolumn{8}{c}{Model 5}\\
$(\phi_1,\phi_2)$ &$N$&{$\infty$} &
$0.45$&{ $0.55$} &{$0.65$}& {KPSS}& {HR} &$(\phi_1,\phi_2)$ & $N$&{$\infty$} &
$0.45$&{ $0.55$} &{$0.65$}& {KPSS} & {HR}\\

        (0.8, 0.3)& 40 &6.2&6.2&6.2&6.2&7.6&0.9   &(0.8, 0.3)&40&11.8&94.3&88.8&82.3&82.0&99.4\\
                         & 70& 6.4&6.4&6.4&6.4&6.2& 0.4 &&70&11.8&96.6&92.7&88.3&90.1& 100 \\
                         &100 & 7.2&7.2&7.2&7.2&7.0& 0.4 &&100&12.1&98.4&95.4&91.8&95.3&100 \\
                (0.9, 0.5)& 40 &6.7&6.7&6.7&6.7&8.5& 0.4 &(0.9, 0.5)&40&11.8&95.3&90.0&84.2&83.5& 99.8 \\
                              &70& 6.5&6.5&6.5&6.5&8.1& 0.0&&70&12.2&97.2&93.8&89.8&89.2& 100 \\
                              &100& 5.6&5.6&5.6&5.6&7.4& 0.0 &&100&11.6&98.6&96.4&92.7&94.8& 100 \\
                $(0.95, 0.9)$& 40 &7.2&7.2&7.2&7.2&9.0& 0.0  &$(0.95, 0.9)$&40&13.1&95.0&90.0&83.9&83.0& 99.6 \\
                                      &70& 7.1&7.1&7.1&7.1&7.3& 0.2 &&70&11.6&97.3&93.8&89.7&90.2& 100 \\
                                      &100& 5.5&5.5&5.5&5.5&8.1& 0.0 &&100&13.7&99.0&96.4&92.3&95.2& 100\\

\multicolumn{8}{c}{Model 3}& \multicolumn{8}{c}{Model 6}\\
 $(\rho_1,\rho_2)$ &$N$&{$\infty$} &
$0.45$&{ $0.55$} &{$0.65$}& {KPSS}& {HR} &$(\rho_1,\rho_2)$& $N$ &{$\infty$} &
$0.45$&{ $0.55$} &{$0.65$}& {KPSS} & {HR}\\

  (0.4, 0.2)&40   & 7.2&8.2&7.4&7.3&22.5&4.8   & (0.4, 0.2)&40&14.8&98.0&95.2&90.6&85.9& 29.2   \\
                  &70& 7.7&7.7&7.7&7.7&17.3& 5.0  &&70&15.4&99.1&97.0&93.8&92.0& 43.4 \\
                  &100& 7.2&7.2&7.2&7.2&18.0& 5.1  &&100&16.6&99.6&98.8&96.5&96.5& 54.5  \\
  (0.5, 0.1)& 40   &8.5&8.9&8.5&8.5&19.6&5.4   &(0.5, 0.1)&40&14.2&99.1&95.9&91.3&84.7& 30.2  \\
                  &70& 8.0&8.0&8.0&8.0&16.6& 6.2  &&70&14.8&99.4&97.2&94.0&91.2& 47.8   \\
                  &100&6.3&6.3&6.3&6.3&17.4& 5.9  &&100&15.0&99.6&98.5&96.2&95.5&60.9  \\
  (0.6, 0.1)& 40   &8.5&12.7&9.6&8.7&26.2&6.0    &(0.6, 0.1)&40&14.5&99.2&97.1&93.3&87.2&27.6  \\
                  &70& 7.3&7.3&7.3&7.3&22.4& 6.8   &&70&15.7&99.7&98.5&96.2&93.5& 37.3 \\
                  &100& 7.6&7.6&7.6&7.6&20.3& 7.0  & &100&16.4&99.8&99.1&97.7&95.7& 44.0 \\
     
\multicolumn{8}{c}{Model 7}& \multicolumn{8}{c}{Model 7}\\
$(\phi_1,\phi_2)$ &$N$&{$\infty$} &
$0.45$&{ $0.55$} &{$0.65$}& {KPSS}& {HR} &$(\phi_1,\phi_2)$& $N$ &{$\infty$} &
$0.45$&{ $0.55$} &{$0.65$}& {KPSS} & {HR}\\
  
  (0.8, 0.3)&40   &6.7&100.0&100.0&99.9&98.5&100.0&(0.9, 0.5)&40&7.0&100.0&100.0&100.0&98.4&100.0 \\
              &70&6.3&100.0&100.0&100.0&99.7&100.0&&70&5.5&100.0&100.0&100.0&99.5&100.0 \\
             &100&7.0&100.0&100.0&100.0&99.8&100.0&&100& 5.9&100.0&100.0&100.0&99.9&100.0\\
  (0.95, 0.9)& 40   & 8.0&100.0&100.0&100.0&98.5&100.0\cr
  &70& 7.3&100.0&100.0&100.0&99.2&100.0\cr
  &100& 6.1&100.0&100.0&100.0&99.9&100.0\cr

\end{tabular}
}
\end{table}

\section*{Acknowledgement}
We thank the editor, associate editor and referees for their constructive comments. Chang and Cheng were supported by the National Natural Science Foundation of China. Chang was also supported by the Center of Statistical Research and the Joint Lab of Data Science and Business Intelligence at Southwestern University of Finance and Economics. Yao was supported in part by an EPSRC grant.
\vspace{-10pt}

\section*{Supplementary material}
Supplementary material available at Biometrika online includes all the technical proofs and  some additional numerical results.


\vspace{-10pt}

\begin{thebibliography}{}
\bibitem[Andrews(1991)]{Andrews1991}
Andrews, D. W. K. (1991). Heteroskedasticity and autocorrelation consistent covariance matrix estimation.
{\sl Econometrica}, {\bf59}, 817--858.




\bibitem[Cavaliere \& Taylor(2007)]{Cavaliere2007}
Cavaliere, G. \&  Taylor, A. M. R. (2007). Testing for unit roots in time
series models with non-stationary volatility. {\sl Journal of Econometrics},
{\bf 140}, 919--947.




\bibitem[Chan \& Wei(1988)]{Chan1988}
Chan, N. H. \& Wei, C. Z. (1988).
Limiting distributions of least squares estimates of unstable
autoregressive processes.  {\sl The Annals of Statistics}, {\bf16}, 367--401.





\bibitem[Chang et al.(2017)]{Changetal2017}
Chang, J., Zheng, C., Zhou, W.-X. \& Zhou, W. (2017). Simulation-based hypothesis testing of high dimensional means under covariance heterogeneity. {\sl Biometrics}, {\bf 73}, 1300--1310.





\bibitem[Dickey \& Fuller(1979)]{DF1979}
Dickey, D. A. \& Fuller, W. A. (1979). Distribution of the estimators for autoregressive
time series with a unit root. {\sl Journal of the American Statistical
Association}, {\bf 74}, 427--431.


\bibitem[Dickey \& Fuller(1981)]{DF1981}
Dickey, D. A. \& Fuller, W. A. (1981). Likelihood ratio statistics for
autoregressive time series with a unit root. {\sl Econometrica}, {\bf 49}, 1057--1072.

\bibitem[Elliott et al.(1996)]{Elliot1996}
Elliott, G., Rothenberg, T. J. \& Stock, J. H. (1996).
Efficient tests for an autoregressive unit root. {\sl Econometrica}, {\bf 64},
813--836.



\bibitem[Hatanaka(1996)]{Hatanaka1996}
Hatanaka, M. (1996). {\sl Time-Series-Based Econometrics: Unit Roots and
Cointegration}. Oxford University Press.


\bibitem[Hualde \& Robinson(2011)]{HualderRobinson2011}
Hualde, J. \& Robinson, P. M. (2011). Gaussian pseudo-maximum likelihood estimation of fractional time series models. {\sl The Annals of Statistics}, {\bf 39}, 3152--3181.


\bibitem[Hylleberg et al.(1990)]{Hylleberg1990}
Hylleberg, S.,  Engle, R. F., Granger, C. F. J. \& Yoo, S. (1990).
Seasonal integration and cointegration. {\sl Journal of Econometrics},
{\bf 44}, 215--238.


\bibitem[Kwiatkowski et al.(1992)]{kpss1992}
Kwiatkowski, D., Phillips, P. C. B., Schmidt, P. \&
Shin, Y. (1992).  Testing the null hypothesis of stationarity
against the alternative of a unit root. {\sl Journal of Econometrics},
{\bf 54}, 159--178.





\bibitem[Maddala \& Kim(1998)]{Maddala1998}
Maddala, G. S. \& Kim, I.-M. (1998). {\sl Unit roots, Cointegration and Structural
Change}. Cambridge University Press.

\bibitem[Nelson \& Plosser(1982)]{np82}
Nelson, C. R. \& Plosser, C. I. (1982). Trends versus random walks in macroeconomic
time series: some evidence and implications. {\sl Journal of Monetary Economics},
{\bf 10}, 139--162.


\bibitem[Paparoditis \& Politis(2005)]{Paparoditis2005}
Paparoditis, E. \& Politis, D. N. (2005).
Bootstrapping unit root tests for autoregressive
time series. {\sl Journal of the American Statistical Association},
{\bf 100}, 545--553.





\bibitem[Pesaran(2007)]{Pesaran2007}
Pesaran, M. H. (2007). A simple panel unit root test in the presence of
cross-section dependence. {\sl Applied Econometrics}, {\bf 22}, 265-317.


\bibitem[Phillips(1987)]{Phillips1987}
Phillips, P. C. B. (1987). Time series regression with a unit root.
{\sl Econometrica}, {\bf 55}, 277--301.




\bibitem[Phillips \& Perron(1988)]{Phillips1988}
Phillips, P. C. B. \& Perron, P. (1988). Testing for a unit root in time
series regression.  {\sl Biometrika}, {\bf 75},  335--346.



\bibitem[Phillips \& Xiao(1998)]{PhillipsXiao1998}
Phillips, P. C. B. \& Xiao, Z. (1998).
A primer on unit root testing. {\sl Journal of Economic Surveys}, {\bf 12}, 423--469.


\bibitem[Rho \& Shao(2019)]{RhoShao2019}
Rho, R. \& Shao, X. (2019). Bootstrap-assisted unit root testing with
piecewise locally stationary errors. {\sl Econometric Theory}, {\bf35}, 143--166.

\bibitem[Robinson(1994)]{Robinson1994}
Robinson, P. M. (1994). Efficient tests of nonstationary hypothesis.
{\sl J. 
Am. Statist. Assoc.}, {\bf 89}, 1420--1437.

\bibitem[Said \& Dickey(1984)]{Said1984}
Said, S. E. \& Dickey, D. A. (1984). Testing for unit roots in autoregressive-moving
average models of unknown order. {\sl Biometrika}, {\bf71}, 599--608.



\bibitem[Stock(1994)]{Stock1994}
Stock, J. H. (1994). Unit roots, structural breaks and trends.
In R. F. Engle and D.  L. McFadden (eds), {\sl Handbook of Econometrics},
Vol.4, Elsevier Science.



%




\bibitem[Zivot \& Andrews(1992)]{Zivot1992}
Zivot, E. \& Andrews, D. W. K. (1992). Further evidence on the great
crash, the oil price shock, and the unit root hypothesis. {\sl Journal of
Business $\&$ Economic Statistics}, {\bf 10},
251--270.

\end{thebibliography}
\end{document}